\documentclass[twocolumn]{article}

\usepackage[english]{babel}

\usepackage[a4paper,top=2cm,bottom=2cm,left=2cm,right=2cm,marginparwidth=1.75cm]{geometry}

\usepackage{natbib}
\usepackage{amsmath}
\usepackage{graphicx}
\usepackage[colorlinks=true, allcolors=blue]{hyperref}

\title{\textbf{Accuracy nudges are not effective against non-harmful deepfakes}}
\author{Juan Jose Rojas-Constain\thanks{Independent Researcher. Email: \href{mailto:jrojasconstain@gmail.com}{jrojasconstain@gmail.com}}}

\begin{document}
\maketitle

\begin{abstract}
\textbf{I conducted a preregistered survey experiment (n=525) to assess the effectiveness of ``accuracy nudges" against deepfakes (\href{https://osf.io/69x27}{osf.io/69x17}). The results, based on a sample of Colombian participants, replicated previous findings showing that prompting participants to assess the accuracy of a headline at the beginning of the survey significantly decreased their intention to share fake news.  However, this effect was not significant when applied to a non-harmful AI-generated video.} 
\end{abstract}

\section{Introduction}\label{sec:intro}

Shifting attention to accuracy through nudges has been proposed as a scalable intervention to reduce the spread of misinformation online. These interventions have proven effective in reducing the spread of false information related to health and political issues \citep{pennycook2021a, pennycook2019, athey2023}. However, recent advances in generative artificial intelligence have made these systems capable of creating increasingly credible messages, known as \textit{deepfakes}, raising concerns about their potential use for influence operations \citep{goldstein2023, dufour2024}. \textit{How well do accuracy nudge interventions work on deepfake sharing intentions?} This study aims to answer this question using a simplified version of the experimental method proposed by \citet{pennycook2021b}.

\section{Experimental Design}\label{sec:exp-design}

An accuracy nudge intervention, consisting of a similar evaluation prompt used by \citet{athey2023}, was applied to a random half of the sample at the beginning of a pre-registered survey experiment (See \href{https://osf.io/69x27}{osf.io/69x17}). Participants were presented with a neutral news headline and asked \textit{``[...] In your opinion, is this headline truthful?}" (Nudge stimuli can be found at \nameref{sec:instruments}). This single question has been consistently found to improve sharing discernment of online information, regardless of the content topic \citep{pennycook2022}. Participants in the control group were not exposed to this intervention.

\begin{figure}[h]
\centering
\includegraphics[width=0.95\linewidth]{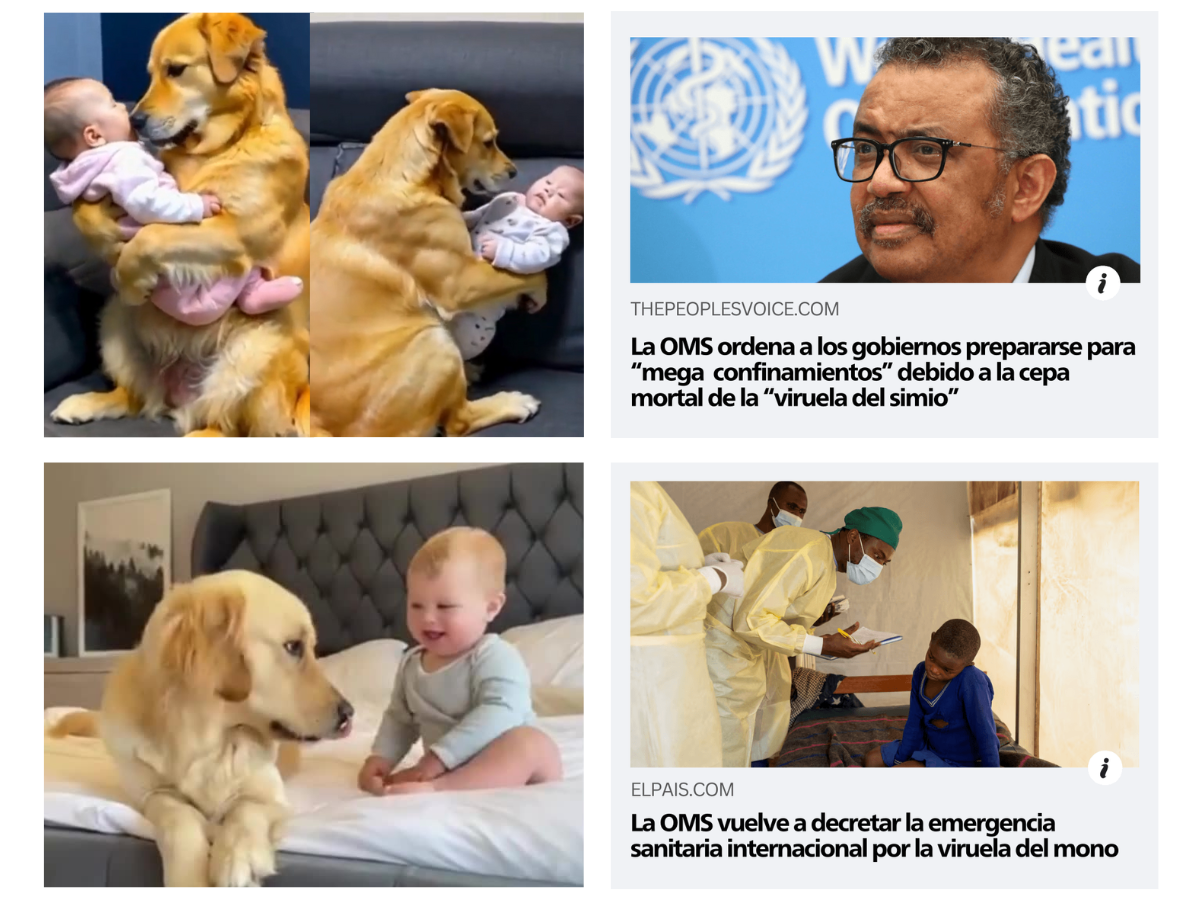}
\caption{\label{fig:stimulus}Stimulus of the content evaluation task. \textit{Top-left:} Deepfake video. \textit{Bottom-left:} Real video. \textit{Top-right:} Real article headlin. \textit{Bottom-right:} False article headline.}
\end{figure}

Subjects were then presented with four types of media content in random order: a deepfake video of a dog carrying a baby like a human, a real video of a dog interacting with a baby, a false, and a real news headline related to the mpox emergency (See \hyperref[fig:stimulus]{figure \ref*{fig:stimulus}}). For each stimulus, participants were asked whether they would share it via private or public online channels (\citealp{athey2023}), to provide a comment of at least 20 characters, and to indicate whether they had encountered the content before.

After the media evaluation task, reflective thinking was measured using three reworded CRT items from \citet{pennycook2019}. Participants reported their political preferences, internet usage (to assess digital literacy), and whether they would ever consider sharing health or cute content online \citep{pennycook2021a}. Education level, age, and gender were also recorded. The survey was implemented using Otree and Heroku, and it was administered online. The complete questionnaire and set of stimuli are available at the \nameref{sec:instruments} section.

\section{Results}

A total of 525 participants completed the survey, with 262 of them (49.90\%) assigned to the treatment group and 261 (49.71\%) identifying as male.  A majority of participants, 291 (55.43\%), reported having more than 16 years of education.  The average age was 32.36 years (SD = 9.67). On a 1-10 scale, the average political position was 4.28 (SD = 1.98), while average internet usage was 4.38 (SD = 0.95) out of 5, and average CRT score was 1.77 (SD = 0.96) out of 3.

\begin{figure}[h]
\centering
\includegraphics[width=1\linewidth]{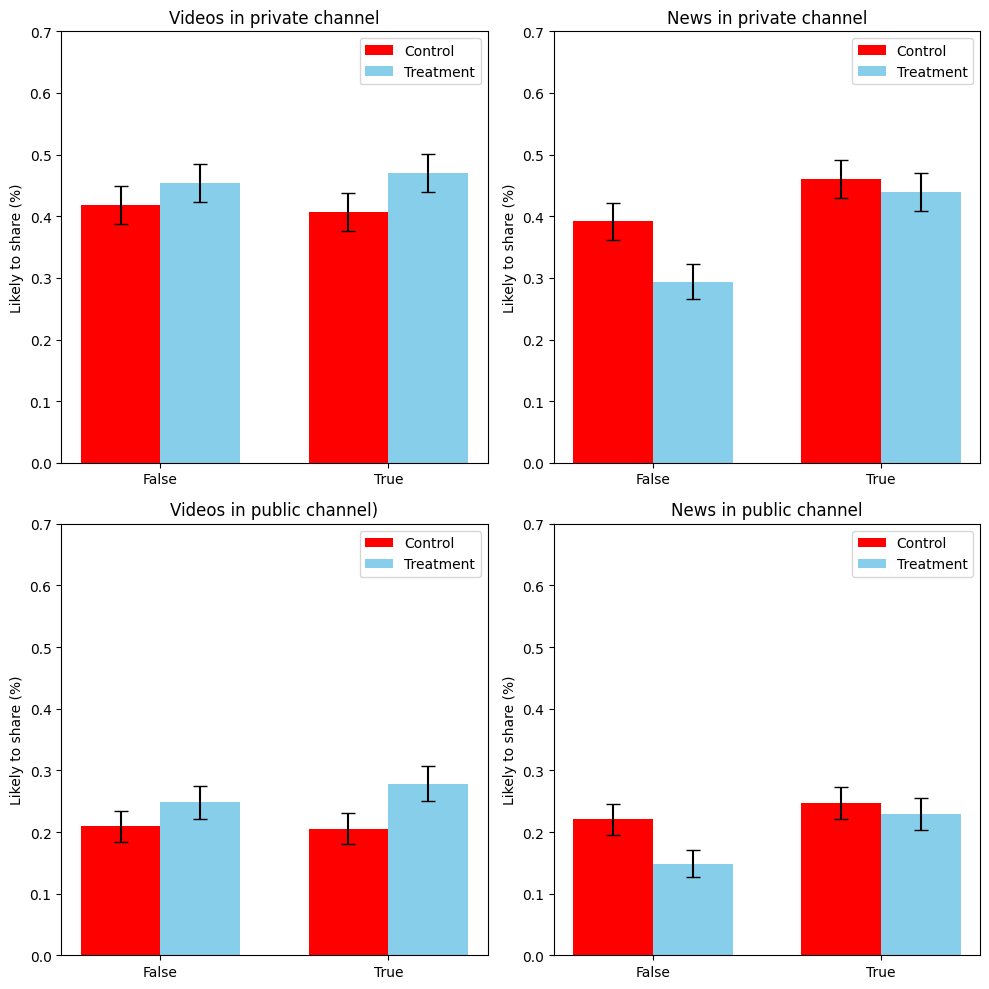}
\caption{\label{fig:results} Likelihood to share content in different channels. Red bars represent the control group, and blue bars represent the treatment group.}
\end{figure}

Fisher's exact tests revealed that participants exposed to the accuracy prompt were significantly less likely to share the fake news headline compared to the control group (\textit{private channels:} Fisher’s exact p = 0.021, 1-sided p = 0.012; \textit{public channels:} Fisher’s exact p = 0.043, 1-sided p = 0.022). However, the proportion of participants willing to share the real news headline was not significantly different between the control and treatment conditions (\textit{private:} Fisher’s exact p = 0.661, 1-sided p = 0.345; \textit{public:} Fisher’s exact p = 0.682, 1-sided p = 0.350) (See top and bottom right plots in \hyperref[fig:results]{figure \ref*{fig:results}}).

Videos tell a different story. Fisher's exact tests showed that the accuracy nudge did not significantly affect participants' intentions to share the deepfake video (\textit{private:} Fisher’s exact p = 0.429, 1-sided p = 0.229; \textit{public:} Fisher’s exact p = 0.300, 1-sided p = 0.169). Similarly, there was no significant effect of the nudge on sharing the real video (\textit{private:} Fisher’s exact p = 0.160, 1-sided p = 0.087; \textit{public:} Fisher’s exact p = 0.053, 1-sided p = 0.031). However, although these differences were not statistically significant, the direction of the results suggests that the accuracy nudge may have slightly increased participants' intentions to share both true and false videos. This trend was observed across private and public channels, indicating that the nudge might not have the same effect on non-harmful video content as it does on news headlines (See top and bottom left plots in \hyperref[fig:results]{figure \ref*{fig:results}}). 

Notably, people are approximately twice as likely to share almost all types of content through private rather than public channels, highlighting the importance of closed messaging platforms such as WhatsApp, Telegram, Messenger, etc., in the dissemination of (mis)information online.

\section{Discussion and Limitations}

My results, based on Colombian subjects, replicate previous findings that accuracy nudges are an effective strategy for reducing the intention to share fake news headlines online. However, they also suggest that the effects of these nudges on fake videos may differ in nature.

These findings are preliminary and provide an initial exploration of the effects of accuracy nudges on AI-generated media content. In future versions of this paper, I will include regression results analyzing the intention to share each type of content on the treatment condition, with controls for gender, education level, age, cognitive reflection (CRT), internet usage, and political extremism. This analysis will offer a more comprehensive understanding of the factors influencing sharing behavior across different content types.

The main limitation of this study is that the two types of content differ not only in their multimedia format but also in their potential consequences. While sharing false content related to the mpox emergency could be harmful, the deepfake video used in this study is not, and sharing it may not have any negative consequences. Given this, the results could have two possible explanations. On one hand, the visual persuasiveness of the video may reduce the effectiveness of accuracy nudges in revealing that a deepfake is false. On the other hand, because the content is harmless and perceived as cute or funny, participants may not consider it important whether the video is real or fake when deciding to share it.

As a first step in assessing these alternatives, I will infer whether participants believed the content based on the comments they provided. Additionally, I will explore the possibility of re-contacting participants to ask whether they consider it important for funny or cute content to be real when deciding to share it. Furthermore, future experiments should use deepfake videos in the same content category as the fake news headlines (p.e. politics), which would allow the question to be answered more effectively.

In any case, these results raise a broader discussion. Online content has multiple dimensions —its format, message, consequences, channels, and so on—. Since experimental methodology is based on isolating variables, a more granular taxonomy of stimuli and the communication environment would enrich the literature and help refine our understanding of the impact of interventions across different content types and channels.

\section*{Survey Instruments}\label{sec:instruments}
The survey questionnaire is available at this link: \href{https://docs.google.com/spreadsheets/d/1tEjPL0rnRgcQuyMzrCGiEIL0UjfxvwLBs9j4TY6FO8U/edit?usp=sharing}{English Questionnaire}

All of the stimuli used in the experiment are available at this link: \href{https://docs.google.com/spreadsheets/d/1tEjPL0rnRgcQuyMzrCGiEIL0UjfxvwLBs9j4TY6FO8U/edit?usp=sharing}{Evaluation Task Stimuli}

\section*{Acknowledgments}
I am deeply grateful to Juan Felipe Ortiz-Riomalo, Lina María Restrepo-Plaza, Allison Benson, Natalia Pérez, and David Becerra for their thoughtful feedback on the design of this study. Thanks to Jhon Villareal for sharing valuable resources that assisted me with the Otree application development. I am also thankful to my girlfriend, Carmen Coy, for her ever-patient listening and support.

\end{document}